\documentclass[anz]{cupaus}
\usepackage{graphicx}

\usepackage{amsmath,amssymb,eulervm}

\begin{document}

\verso{Noise-induced modulations reduce timing jitter}
\recto{Noise-induced modulations reduce timing jitter}

\title{Influence of noise-induced modulations on the timing stability of passively mode-locked semiconductor laser subject to optical feedback}

\cauthormark 
\author[1]{Lina Jaurigue}
\cauthormark 
\author[1]{Kathy Lüdge}

\address[1]{Institute of Physics, Technische Universität Ilmenau, Weimarer Straße 25, 98693 Ilmenau, Germany. \email[1]{lina.jaurigue@tu-ilmenau.de}}

\address[1]{Institute of Physics, Technische Universität Ilmenau, Weimarer Straße 25, 98693 Ilmenau, Germany. \email[2]{kathy.luedge@tu-ilmenau.de}}
      
\pages{1}{18}
      
\begin{abstract}
We show that passively mode-locked lasers subject to feedback from a single external cavity can exhibit large timing fluctuations on short time scales despite having a relatively small long-term timing jitter. This  means that the commonly used von Linde and K\'{e}f\'{e}lian techniques of experimentally estimating the timing jitter can lead to large errors in the estimation of the arrival time of pulses. We also show that adding a second feedback cavity of the appropriate length can significantly suppress noise-induced modulations that are present in the single feedback system. This reduces the short time scale fluctuations of the interspike interval time and at the same time improves the variance of the 
fluctuation of the pulse arrival times on long time scales. 
\end{abstract}

\keywords[\textit{Mathematics subject classification}]{00A71,00A72,00A79}
\maketitle

\section{Introduction}
Passively mode-locked semiconductor lasers are envisaged as a compact, inexpensive alternative to other pulsed sources of light 
\cite{AVR00,LEL07,LUE12,RAF07}. However, for this to be realized the issue of their relatively large timing jitter needs to be overcome. To this end research has been carried out on various techniques of improving the timing regularity of such devices. These techniques include hybrid mode-locking \cite{FIO10,ARK13}, optical injection \cite{REB10,REB11, DUQ23}, opto-electronic feedback \cite{DRZ13} and optical feedback \cite{SOL93,LIN10e,ARS13,BRE10,DRZ13a,AVR09,OTT12a,OTT14b,JAU15,SIM14,DON20,WAN18d,WAN21b, LI21b, ASG18,ASG17}. All-optical feedback has been of particular interest as this technique is very simple to implement and does not require any additional electronics. Theoretical studies on this topic have predicted that a reduction in the timing jitter can be achieved when resonant feedback lengths are chosen, i.e. integer multiples of the period of the laser \cite{OTT12a,OTT14}, and that better reduction can be achieved for longer delay times \cite{JAU15}. Experimentally a resonance structure in dependence on the delay length has also been observed, as well as improved timing jitter reduction for longer feedback cavities \cite{LIN10e,ARS13,NIK16,ASG17}. 
However, for longer delay times the influence of noise-induced modulations plays an increasingly important role. These modulations
are seen experimentally as side peaks in the power spectra of the laser output and are some times referred to as supermode resonances. So far there appears to be a lack of understanding of how these noise-induced modulations affect the timing regularity of passive 
mode-locking lasers \cite{BRE10,DRZ13a,HAJ12,DON20,WAN21b,WAN18d,ASG18}. In this paper we will show how noise-induced modulations affect the pulse train and how this influences the common measures of the timing jitter.

Noise-induced oscillations (the underlying deterministic system is in a steady state) or modulations (the underlying deterministic system is exhibits oscillatory dynamics) 
have been studied and observed in a wide range of systems \cite{CHE05d,POT02a,SIG89,BAL04,GOL03,JAU16a}. There have also been several works on the suppression thereof, particularly using feedback \cite{FLU07,JAU16a,JAU16,HAJ12,POT02a}.
In this paper we will show how the timing regularity of the mode-locked laser output can be improved by implementing a dual feedback cavity
configuration, where the second cavity acts to suppress the noise-induced modulations arising due to the first feedback cavity.

The paper is structured as follows. In Section \ref{sec:model} and \ref{sec:methods} we present the model for the passively mode-locked laser subject to optical feedback and various methods of determining long-term timing jitter. In Section \ref{sec:SFB}
we look at the influence of resonant optical feedback from one external cavity on the regularity of the pulse train 
and demonstrate the impact of 
noise-induced modulations. In Section \ref{sec:map} we propose a simplified iterative map that can reproduce the effect of the modulations. Subsequently, in Section \ref{sec:DFB} we show that the noise-induced modulations can be suppressed via the addition of a second
feedback cavity with an appropriately chosen length and we identify optimal conditions for the second feedback cavity in order to 
achieve both modulation suppression 
and low long-term timing jitter. Finally, we discuss the results and conclude.

\section{Methods}
\subsection{Delay differential equation model}\label{sec:model}

\begin{figure}[tb]
\centering\includegraphics[width=0.98\columnwidth]{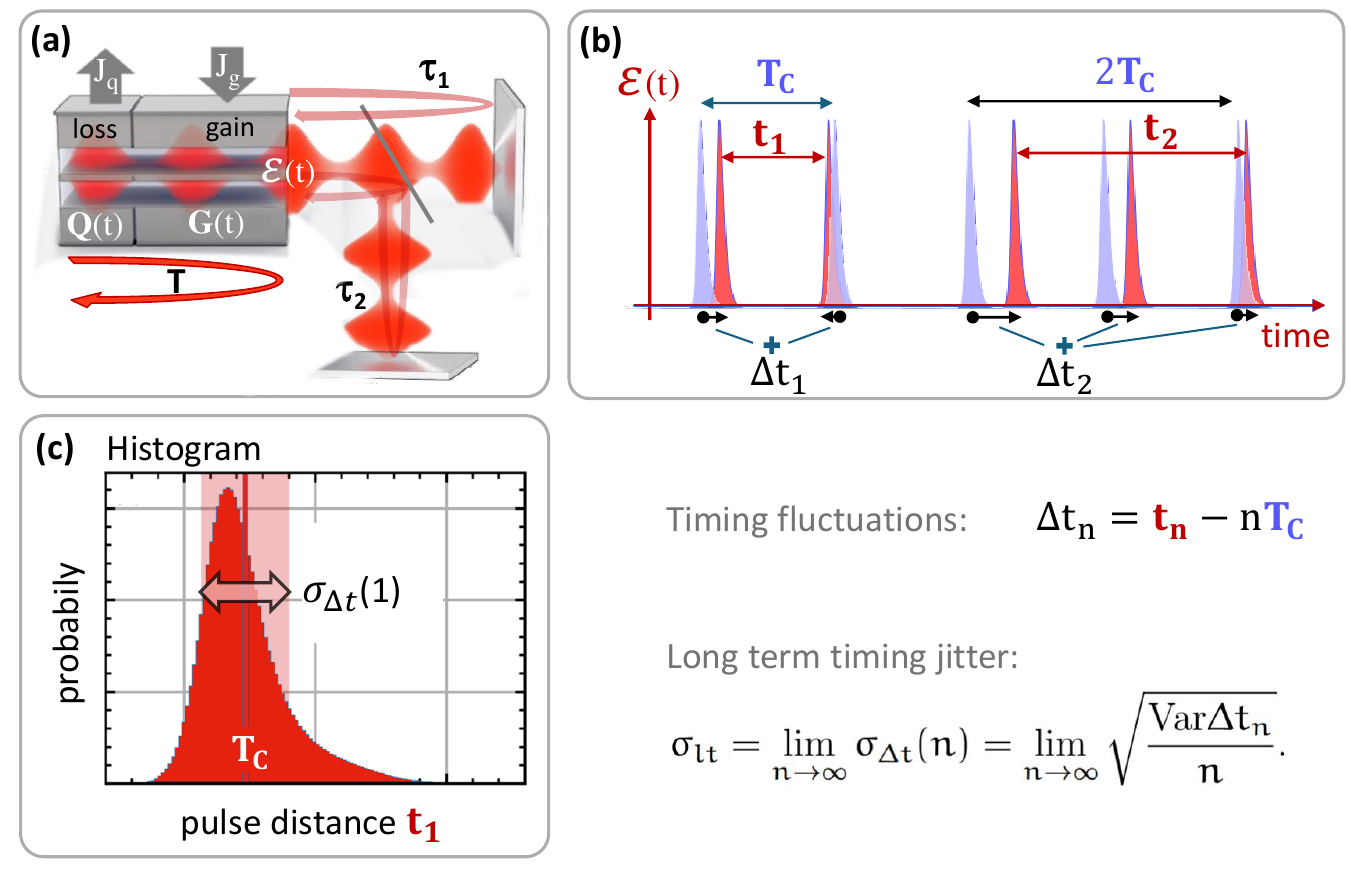}
\caption{(a) Sketch of the two-section passively mode-locked laser with two external feedback cavities explaining the quantities in the delay differential equation model (equations \eqref{subeq:E}-\eqref{subeq:Q}), (b) Sketch of the emitted pulse train in red, blue pulses represent idealized pulses with equal distance $T_C$ ($T_C$ is the mean pulse distance), arrows indicate definition of timing fluctuations $\Delta t_n$, (c) Visualization of the distribution of pulse intervals $t_1$, \textit{pulse-to-pulse timing jitter} is $\sigma_{pp}=\sigma_{\Delta t}(n=1)$, \textit{long term jitter} is defined as  $\sigma_{lt}=\sigma_{\Delta t}(n\rightarrow \infty)$. }\label{fig:model}
\end{figure}

We use the model presented in \cite{JAU16} for a two section passively mode-locked semiconductor laser subject to optical feedback from two 
external cavities (a sketch of the setup is shown in figure\,\ref{fig:model}a). The model is based on that proposed in  \cite{VLA04,VLA05,VLA10}, which was extended to include optical feedback in \cite{OTT12a}.
The laser is described by a set of three delay differential equations:
\begin{eqnarray}
 \qquad \dot{\mathcal{E}}\left(t\right)
&=&-\gamma\mathcal{E}\left(t\right)+\gamma R\left(t-T\right) e^{-i \Delta \Omega T}
\mathcal{E}\left(t-T\right)+\sqrt{R_{sp}}~\xi\left(t\right)\nonumber\\
 &+& \sum_{l=1,2}K_l e^{i l C_l} \gamma R\left(t-T-l\tau_l\right) e^{-i \Delta \Omega
\left(T+l\tau_1\right)}
 \mathcal{E}\left(t-T-l\tau_1\right), ~~~~~\label{subeq:E}\\ ~\nonumber\\
 \dot{G}\left(t\right)&=& J_g-\gamma_g
G\left(t\right)-e^{-Q\left(t\right)}\left(e^{G\left(t\right)}-1\right)|\mathcal{E}\left(t\right)|^2,
\label{subeq:G}\\
 \dot{Q}\left(t\right)&=&J_q-\gamma_q Q\left(t\right)-r_s
e^{-Q\left(t\right)}\left(e^{Q\left(t\right)}-1\right)|\mathcal{E}\left(t\right)|^2,\label{subeq:Q}
\end{eqnarray}
with
\begin{equation}
 R\left(t\right)\equiv\sqrt{\kappa} e^{\frac{1}{2}\left(\left(1-i
\alpha_g\right)G\left(t\right)-\left(1-i \alpha_q\right)Q\left(t\right)\right)} \label{defR}
\end{equation}
being the change of the slowly varying electric field amplitude
$\mathcal{E}$ during one roundtrip in the ring cavity ($\alpha_g$, $\alpha_q$ are the linewidth enhancement factors in the gain and absorber sections which we  assume to be zero \cite{JAU17a}). All transmission, coupling and internal losses at the interfaces between the different
sections are lumped together into the attenuation factor $\kappa$.

The saturable gain $G$ and the saturable loss $Q$ are effective dynamic variables that describe dimensionless carrier densities. They have been obtained from the traveling wave model via integration of the carrier densities within the gain and absorber sections \cite{OTT12a}. The same holds true for the pump parameters $J_g$ and $J_q$ which are integrated and rescaled values that are proportional to the injection current in the gain section and to reverse bias induced carrier losses in the absorber section \cite{OTT12a}. The carrier decay rates inside the gain and absorber sections are denoted 
by $\gamma_g$ and $\gamma_q$, respectively (for semiconductor materials the lifetime of carriers within the gain $\gamma_g$ is much smaller than in the absorber $\gamma_q$ \cite{AGR93}).

The finite width of the gain spectrum is taken into account by a bandwidth-limiting element,
which is described by a Lorentzian-shaped filter function with a width of $\gamma$. If the gain is centered at the optical frequency of the closest cavity mode (assumed here for simplicity) the detuning $\Delta\Omega$ is zero. Inhomogeneous broadening is not taken into account here but could also be included in the model \cite{PIM19}.
The last terms in equations \eqref{subeq:G} and \eqref{subeq:Q} describe the light–matter
interactions which leads to a depletion by the pulse that travels in the cavity.
The factor $r_s$ is proportional to the ratio of the linear gain coefficients of the absorber and gain
sections. Spontaneous emission noise is taken into account in equation \eqref{subeq:E} via a complex Gaussian white noise term $\xi(t)$ with a strength $\sqrt{R_{sp}}$ where $R_{sp}$ is the spontaneous emission rate. 
 All laser parameters are as described in \cite{JAU16} and given in Table \ref{tab:fonts} and correspond to parameters that model the experimentally observed emission of  semiconductor-based two-section passively mode-locked lasers \cite{NIK16}.

$K_1$ and $K_2$ are the feedback strengths from the two external cavities with delay lengths $\tau_1$ and $\tau_2$, respectively.
Here, under the assumption of weak feedback we only include contributions to the feedback from light that has made one roundtrip 
in the feedback cavities \cite{OTT12a}. $C_1$ and $C_2$ are the phase shifts accumulated after one roundtrip in the external cavities.

Feedback influences the dynamics of the mode-locked laser \cite{OTT12,OTT14}, however in this paper
we are primarily interested in the reduction of the timing jitter for fundamental mode-locking, we therefore restrict our study to only resonant 
feedback delay times, i.e integer multiples of the mode-locked pulse period which we denote as $T_0$. Note that due to the light matter interaction within the cavity, the period of the deterministic solution $T_0$ is slightly larger than the cold cavity round-trip time $T$. Additionally, we set the feedback phases $C_l$ to zero, as these have no qualitative 
influence on the timing jitter reduction trends in dependence of the feedback delay lengths \cite{OTT12a,JAU16}. When we discuss
single cavity feedback ($K_2=0$), we will refer to the feedback strength $K$ and delay length $\tau$.
In the following, the parameter values used will be those given in Table 1, unless stated otherwise. 
The laser cavity roundtrip used here, $T=25$\,ps, corresponds to a 1\,mm Fabry-Perot cavity \cite{OTT12}.

\begin{table}[h]
\caption{Parameter values used in numerical simulations, unless stated otherwise.}
\label{tab:fonts}
\begin{center}
\begin{tabular}{|l|l|l|l|l|l|l|} 
\hline
\rule[-1ex]{0pt}{3.5ex}  Parameter & Value &  Parameter & Value &  Parameter & Value\\
\hline
\hline
\rule[-1ex]{0pt}{3.5ex} $\gamma$&2.66\,ps$^{-1}$&$\kappa$& 0.1& $T$ & 25\,ps \\
\hline
\rule[-1ex]{0pt}{3.5ex}  $\gamma_g$& 1\,ns$^{-1}$&$\alpha_g$ &  0& $R_{sp}$& 0.04\,ps$^{-1}$ \\
\hline
\rule[-1ex]{0pt}{3.5ex} $\gamma_q$&75\,ns$^{-1}$& $\alpha_q$ & 0 &$C_1$ & 0 \\
\hline
\rule[-1ex]{0pt}{3.5ex} $J_g$& 0.12\,ps$^{-1}$&$r_s$&25& $C_2$ & 0  \\
\hline
\rule[-1ex]{0pt}{3.5ex} $J_q$ & 0.3\,ps$^{-1}$& $\Delta\Omega$ &  0 &$T_0$&25.373\,ps\\
\hline
\end{tabular}
\end{center}
\end{table}

\subsection{Timing jitter calculation methods}\label{sec:methods}

For passively mode-locked lasers there are several measures of the timing jitter that are commonly used. Experimentally
the so called root mean square (rms) timing jitter is usually calculated using the von Linde method \cite{LIN86}, which involves integrating over the side band of a peak in the power spectrum of the laser output, or the  timing jitter is calculated via the K\'{e}f\'{e}lian method from the width of the fundamental peak in the power spectrum \cite{KEF08}. This is commonly used in experimental studies \cite{KEF08,DRZ13,DRZ13a,DON20,WAN21b}. For our numerical calculations  it is more convenient to derive the \textit{long-term timing jitter} directly from the timing fluctuations \cite{LEE02,OTT14b}. Please see figure\ref{fig:model}b for a visualization of their definition. We will compare this approach also to the power spectrum based K\'{e}f\'{e}lian method. 
In the following, we present the three methods that are used in this study to calculate the timing jitter: 

1) The K\'{e}f\'{e}lian method \cite{KEF08}  assumes a Lorentzian line shape of the peaks in the power spectrum and gives the \textit{long-term timing jitter} $\sigma_{lt}(h)$ as a function of the number of the harmonic $h$  via:
\begin{equation}\label{eq:Kef}
\sigma_{lt}(h)=T_C\sqrt{\frac{\Delta \nu_h T_C}{2\pi h^2}},
\end{equation}
where $\Delta \nu_h$ is the width of the Lorentzian fitted to the $h^\textrm{th}$ harmonic of the power spectral density of the laser output $| E|^2$ and $T_C$ is the mean pulse period. Note that the quantity defined by equation\,\eqref{eq:Kef} is also sometimes referred to as the \textit{pulse-to-pulse jitter} \cite{KEF08,DRZ13a} which, however, rather  describes the width of the inter-pulse distance distribution (see figure\ref{fig:model}c for a visualization).

2) The timing jitter can also be calculated directly from the pulse arrival times \cite{LEE02,OTT14b}. Defining timing fluctuations as shown in figure\ref{fig:model}b via
\begin{equation}
    \Delta t_n \equiv t_n - nT_C,
\end{equation}
where $t_n$ is the arrival time of the $n^\textrm{th}$ pulse and $nT_C $ is the "ideal" arrival time of the $n^\textrm{th}$ pulse in the jitter free case, the \textit{long-term timing jitter} is given by:
\begin{equation}\label{eq:lttj}
\sigma_{lt}=\lim_{n \rightarrow \infty} \sigma_{\Delta t}(n)=\lim_{n \rightarrow \infty}\sqrt{\frac{\textrm{Var}\Delta t_n}{n}}.
\end{equation}
As a passively mode-locked laser does not have a clock time, $T_C$  
is the average interspike interval time calculated over many noise realisations (experimental runs).  We calculate the timing fluctuations and the variance Var${\Delta t_n}$ thereof via detecting the pulse positions $t_n$ in a long train of emitted pulses as described in \cite{OTT14b}. 
It is noted  that there are also other very efficient methods for timing jitter calculation as published by Meinecke et al. \cite{MEI21}, however for our case with side modes they are not suitable.

3) For numerical simulations of passively mode-locked lasers with optical feedback there is also a semi-analytic method of calculating the \textit{long term timing jitter} \cite{PIM14b,JAU15}. In \cite{JAU15} the following expression was derived from the semi-analytic timing-jitter for for resonant feedback delay lengths $\tau=q T_{C}$, $q\in \mathbb{N} $:
\begin{equation}
 \sigma_{lt}(q)=\frac{\sigma_{lt}^{\tau=0}}{1+qK\mathcal{F}\left(K\right)}. \label{semi}
\end{equation}
Here $\sigma_{lt}^{\tau=0}$ is the timing jitter for instantaneous feedback ($\tau=0$) and 
$\mathcal{F}\left(K\right)$ semi-analytic expression depending on the laser parameters and feedback strength and must be calculated numerically, as described in \cite{JAU15}. Equation \eqref{semi} holds for sufficiently small noise terms in equations\,\eqref{subeq:E}-\eqref{subeq:Q} and as long as all non-neutral eigenmodes of  equations\,\eqref{subeq:E}-\eqref{subeq:Q}, linearised about the mode-locked solution, have eigenvalues $\lambda$ with $\textrm{Re}\lambda\ll 0$, i.e. as long as
the influence of noise-induced modulation can be neglected.

\section{Results}
\subsection{Timing jitter in passively mode-locked lasers subject to optical feedback from one external cavity}\label{sec:SFB}

In the limit that the timing fluctuations behave like a random walk all the measures of the timing jitter described in Section\,\ref{sec:methods} are directly proportional to
one another \cite{OTT14b,JAU15}. For a solitary laser this limit is reached when one considers the timing fluctuations over time spans that are much longer than 
any internal time scales that exist in the laser system. The solitary laser pulse positions are correlated over a few roundtrips in the laser cavity due
to the finite time the system requires to recover from perturbations and thus the internal time scale is determined by the relaxation 
oscillation frequency. This means
that on time scales of a few thousand laser cavity roundtrips the timing fluctuations behave like a random walk, meaning that the variance of the timing fluctuations 
grows linearly with the number of roundtrips \cite{OTT14b,JAU15}.

\begin{figure}[t]
\centering\includegraphics[width=0.99\textwidth]{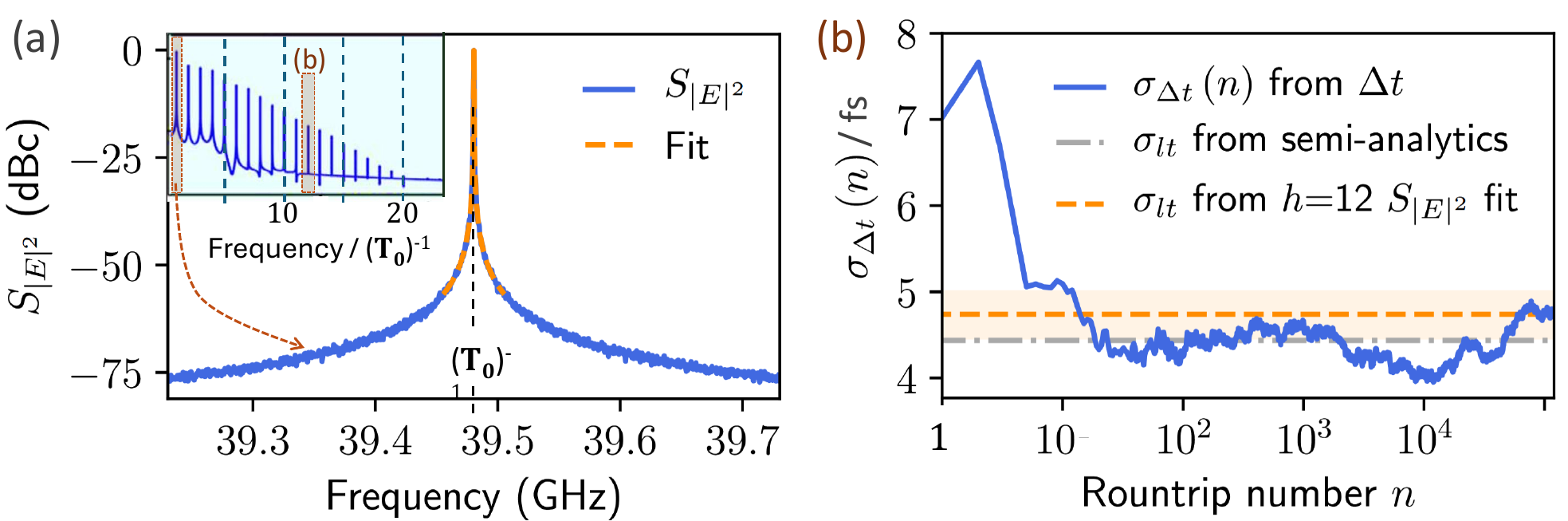}
\caption{\textbf{Spectra and timing jitter without delay}:(a) Zoom of first peak of the electric field power spectrum $S_{|E|^2}$ (blue) with a Lorentzian line-shape fitted to the first harmonic peak (orange, dashed), inset shows full power spectrum, $T_0$ is the mode-locking period. (b) Normalised standard deviation of the pulse arrival times $\sigma_{\Delta t}(n)$ (equation \eqref{eq:tj}) as a function laser cavity round-trip number (blue). Grey dash-dotted and orange dashed line depict timing jitter obtained via semi-analytic method (equation \eqref{semi} with $\sigma_{lt}^{\tau=0}=4.4$\,fs, $\mathcal{F}(K)=0.9$) and K\'{e}f\'{e}lian method (equation \eqref{eq:Kef}, $h=12$, error range shaded in orange) respectively.}\label{fig:sol}
\end{figure}

By adding optical feedback to a passively mode-locked laser an additional time scale is introduced to the system, 
which has two major effects. Firstly, the pulse positions become correlated over the delay time of the feedback cavity which leads to a reduction in
the timing jitter. This was shown in \cite{JAU15} using a semi-analytic approach to calculating the long-term timing jitter. Secondly, the stability of
the mode-locked solution is decreased which can lead to noise-induced modulations with frequencies related to the feedback delay time \cite{JAU16a,YAN09}.
We will now look at the consequences of these two effects.

\begin{figure}[t]
\centering\includegraphics[width=0.99\textwidth]{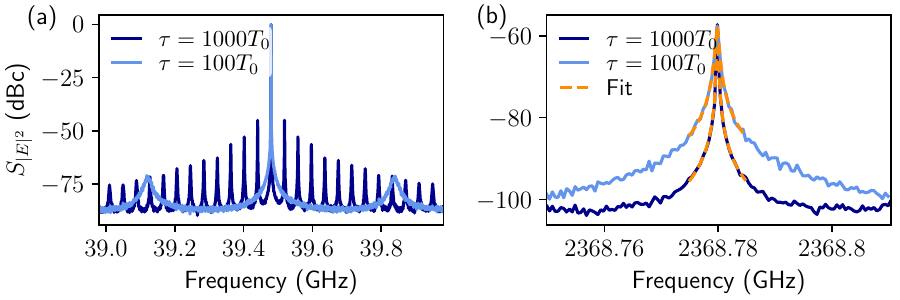}
\caption{\textbf{Delay induced side modes:} Power spectra $S_{|E|^2}$ of the electric field intensity for delay values of $\tau=100 T_0$ (light blue) and $\tau=1000 T_0$ (dark blue). (a) zoom around 1$^{st}$ harmonics, (b) zoom around the central peak of the $60^{th}$ harmonics with the Lorentzian fit indicated in orange. }\label{taufft}
\end{figure}

In figure\,\ref{fig:sol}a the power spectrum of the simulated mode-locked laser without feedback is shown at the fundamental frequency ($1^{st}$ harmonic) while the inset shows the full spectrum. Figure\,\ref{fig:sol}b shows the timing jitter $\sigma_{\Delta t}(n)$ determined from the timing fluctuations $\Delta t_n$ as a function of the roundtrip number $n$ (defined in equation\,\eqref{eq:tj}).
\begin{equation}\label{eq:tj}
\sigma_{\Delta t}(n)=\sqrt{\frac{\textrm{Var}\Delta t_n}{n}}.
\end{equation}
For $n$ much longer than the timescales of the dynamics of the mode-locked laser the jitter $\sigma_{\Delta t}$(n) determined via equation\,\eqref{eq:tj} converges to the long-term jitter given by equation\,\eqref{eq:lttj} and there is an agreement with all three methods used here, i.e. the timing fluctuations described by equation\,\eqref{eq:lttj} (blue line in figure\,\ref{fig:sol}b) the \textit{long-term timing jitter} calculated semi-analytically using equation\,\eqref{semi} (grey dot-dashed line), the \textit{long-term timing jitter} described by the  K\'{e}f\'{e}lian method and calculated using equation\,\eqref{eq:Kef} with $h=12$ (orange dashed line).


When resonant feedback is introduced, the linewidth of the fundamental peak is reduced and supermodes start to appear in the power spectrum (the frequency spacing is given by $\approx 1/\tau$). Examples of experimentally observed noise peaks are found in \cite{BRE10,SOL93,ARS13}. Figure\,\ref{taufft}a shows example spectra for single cavity feedback with delay times of $\tau=100T_0$ and $\tau=1000T_0$, where $T_0$ is the period of the underlying deterministic system. We only consider resonant feedback and thus only delay times that are integer multiples of the natural pulse repetition frequency $1/T_0$. For non-resonant feedback conditions, complex emission dynamics is observed that is however not the focus of this work and studied in more detail elsewhere \cite{JAU17,LUE19, SEI22}. 

\begin{figure}[t]
\centering\includegraphics[width=12cm]{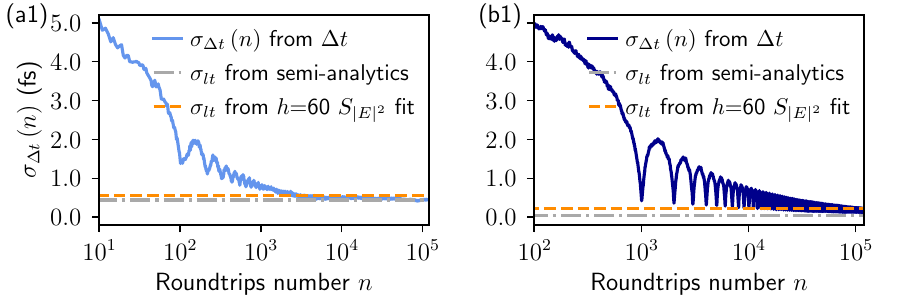}
\centering\includegraphics[width=12cm]{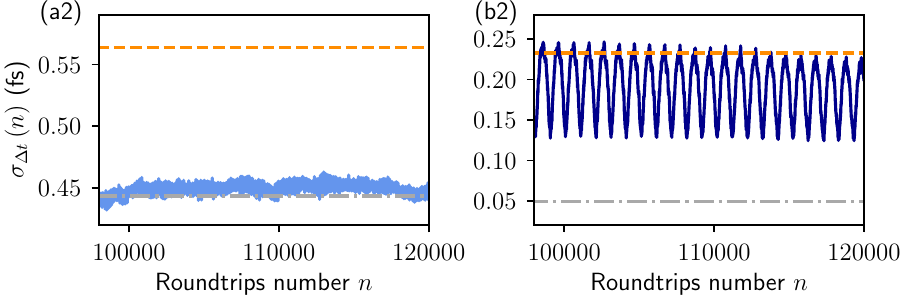}
\caption{\textbf{Timing jitter of mode-locked laser with one delay}: Normalised standard deviation of the pulse arrival times $\sigma_{\Delta t}(n)$ as a function of the laser cavity round-trip number n (blue lines). Grey and orange dotted lines indicate timing jitter obtained via the semi-analytic method ($\sigma_{lt}^{\tau=0}=4.4$\,fs, $\mathcal{F}(K)=0.9$) and the K\'{e}f\'{e}lian method ($h=60$) analogous to figure\ref{fig:sol} for two different delay values (a): $\tau=100 T_0$, and (b): $\tau=1000 T_0$. Panels (a2) and (b2) are zooms of (a1) and (b1).  Parameters: $K=0.1$, all other parameters as in Table \ref{tab:fonts}.}\label{tau_jitter}
\end{figure}

The timing jitter $\sigma_{\Delta t}(n)$ that results from the timing fluctuations  is depicted in figure\,\ref{tau_jitter} with blue lines. The long-term timing jitter calculated via the K\'{e}f\'{e}lian method (equation\,\eqref{eq:Kef} and orange dashed lines in figure\,\ref{tau_jitter}) was obtain by fitting the $60^\textrm{th}$ harmonic of the power spectrum (figure\,\ref{taufft}b). Due to the linewidth reduction with the resonant feedback, it was necessary to fit the Lorentzian to a high harmonic such that the frequency resolution allowed for a sufficiently accurate fit. 

Comparing the jitter curves of the two delays in figure\,\ref{tau_jitter} panel (a1) and (b1), it is apparent that for $\tau=100 T_0$ the jitter $\sigma_{\Delta t}$ converges to a constant value in much fewer roundtrips than for $\tau=1000 T_0$. This is a consequence of the pulse positions being correlated over longer times in the longer delay case. Further, the jitter plots for the long delay show clear evidence of noise induced modulation of the pulse arrival times. In the $\tau=1000 T_0$ case in figure\,\ref{tau_jitter}b1, pronounced fluctuations in $\sigma_{\Delta t}$ are present as can be seen by the oscillating blue curve. These fluctuations have a period roughly equal to $\tau$ and are caused by noise-induced modulations (as was also visible in the power spectra in figure\,\ref{taufft}a). Noise-induced modulations are also present for smaller delay times, however the effect is not as pronounced since the damping rate scales with $1/\tau$.  The presence of the fluctuations for $\tau=1000 T_0$ means that $\sigma_{\Delta t}$ never truly converges to a constant value. 

The above observations have several consequences. Firstly, it means that the timing fluctuations are not wide-sense stationary \cite{DRA67}, which means that one can not use the Wiener-Khinchin theorem \cite{GAR04} to relate the autocorrelation function of the timing deviations to the phase noise spectrum \cite{OTT14}. 
Secondly, in a strict sense, the long-term timing jitter can not be calculated reliably when fluctuations are present and  the long-term timing jitter calculated from the power spectrum can not be interpreted as the variance of the timing deviations. Furthermore, the lower limit of the jitter $\sigma_{\Delta t}$ does not reach the semi-analytic limit (dash-dotted grey line in figure\,\ref{tau_jitter}b2). Whereas, for $\tau=100T_0$ random-walk-like behaviour is regained in the jitter $\sigma_{\Delta t}(n)$ after approximately $10^4$ laser cavity roundtrips and the semi-analytic limit (dash-dotted grey line in figure\,\ref{tau_jitter}a2) is reached.

\begin{figure}[t!]
    \begin{center}
        \includegraphics[width=0.95\textwidth]{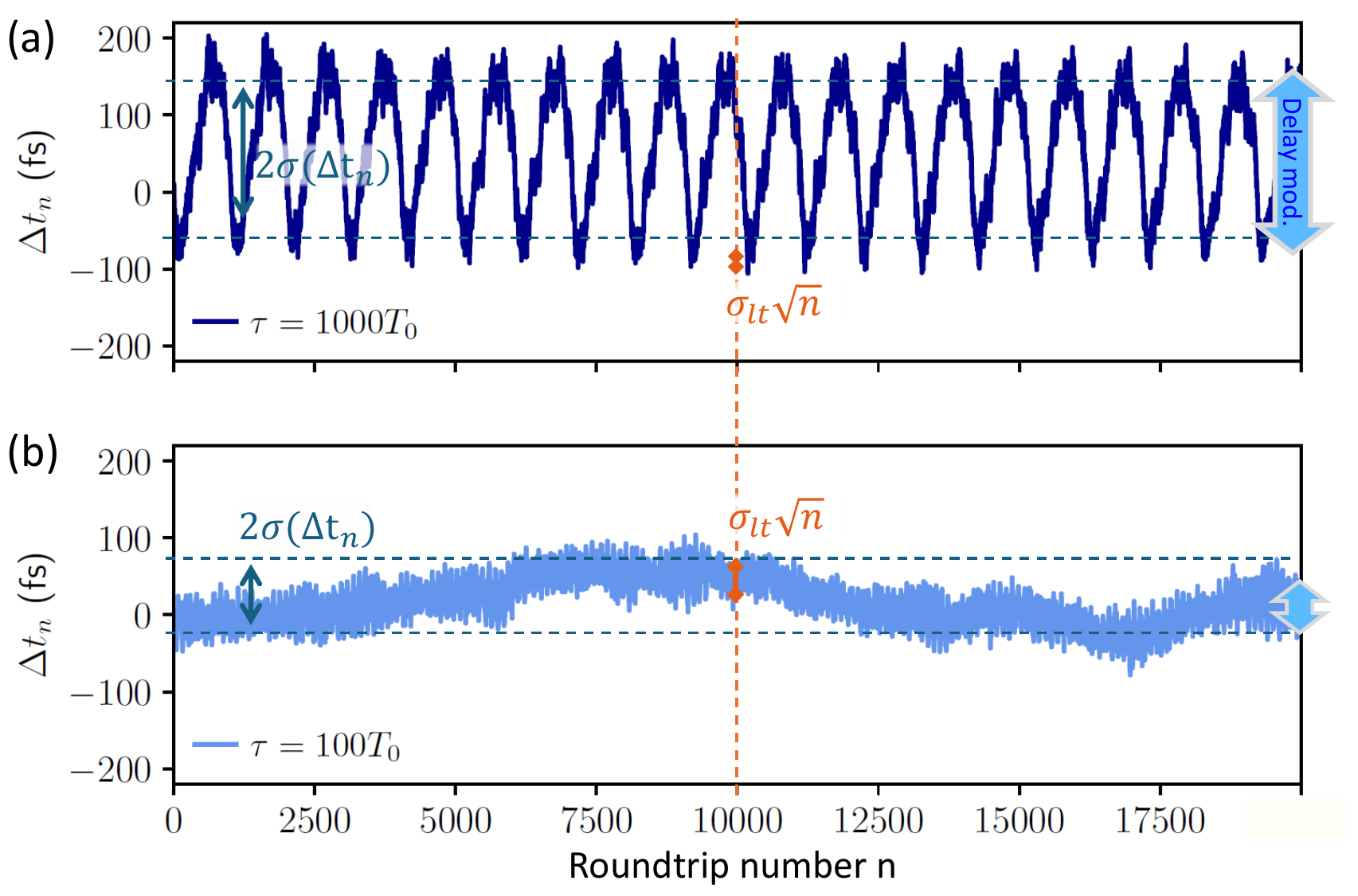}
        \caption{\textbf{Timing fluctuations} $\Delta t_n$ as a function of the round trip number $n$ for  (a)  $\tau=1000T_0$ and (b)  $\tau=100T_0$. Delay induced fluctuations with amplitudes of $\approx 130\,fs$ (blue arrow on the right)  dominate $\sigma(\Delta t_n)_{\tau=1000}=78$fs for $\tau=1000T_0$ while they have smaller amplitudes of  $ \approx 40\,fs$ for $\tau=100T_0$ where stochastic fluctuations dominate $\sigma(\Delta t_n)_{\tau=100}$=30fs. Orange arrows indicate value for $\sigma_{\Delta t}(10^4)$ obtained from long term jitter (13\,fs, 44\,fs) .
        Parameters: $K=0.1$ all other parameters as in Table \ref{tab:fonts}.}\label{t_fluct}
    \end{center}
\end{figure}

We have established that the noise-induced oscillation increase the variance of the timing fluctuations (i.e. the standard deviation of timing fluctuations $\sigma(\Delta t_n)=\sqrt{ Var \Delta t_n}$). In addition to this, the more direct effect is that they cause periodic modulations in the timing fluctuations. An example of this is shown in figure\,\ref{t_fluct} where the timing fluctuations $\Delta t_n$ are depicted for one noise realisation as a function of the roundtrip number for $\tau=1000T_0$ (a) and $\tau=100T_0$ (b). If the amplitude of these delay induced modulations is sufficiently large, then the pulse arrival times can deviate significantly from the expected arrival time interval based on the long-term jitter. To illustrate this we calculate the expected normalized standard deviation of the pulse arrivals times $\sigma(\Delta t_n)$ after $10^4$ laser cavity roundtrips from the long-term timing jitter and compare this with the amplitude of the timing fluctuations depicted in figure\,\ref{t_fluct}. For $\tau=100T_0$, the long-term timing jitter is $\sigma_{lt}\approx0.44$\,fs (see figure\,\ref{tau_jitter}a2). After $10^4$ roundtrips this corresponds to a standard deviation of  $\sigma(\Delta t_n)=\sigma_{lt}\sqrt{n}=44$\,fs, which dominates the amplitude of the timing fluctuations $\Delta t_n$ while the delay induced amplitudes are $\approx 40$\,fs in figure\,\ref{t_fluct}b . To make the same comparison for the $\tau=1000T_0$ case we take the value that the minima of $\sigma_{\Delta t}$ converge to, i.e. a long-term timing jitter of $\sigma_{lt}\approx 0.13$\,fs as seen in figure\,\ref{tau_jitter}b2. In this case the standard deviation after $10^4$ roundtrips is $\sigma(\Delta t_n)=\sigma_{lt}\sqrt{n}=13$\,fs, which is an order of magnitude smaller than the amplitude of the modulation of $\Delta t_n$ shown in figure\,\ref{t_fluct}a ($\approx 130$\,fs).

We can thus conclude that the long-term timing jitter gives an estimate of the underlying random walk-like behaviour, but contains no information on timing deviations on short time scales. For long feedback delay times the amplitude of the noise-induced modulations of the timing deviations can be larger than the underlying drift in the timing fluctuations, even on relatively long time scales ($10^5$ roundtrips corresponds to approximately $2.5$\,ms). Therefore, neglecting the influence of the noise induced oscillation can lead to significant errors in the estimation of pulse arrival times.

\subsection{Timing jitter for a driven iterative map}\label{sec:map}

\begin{figure}[t!]
    \begin{center}
        \includegraphics[width=0.9\textwidth]{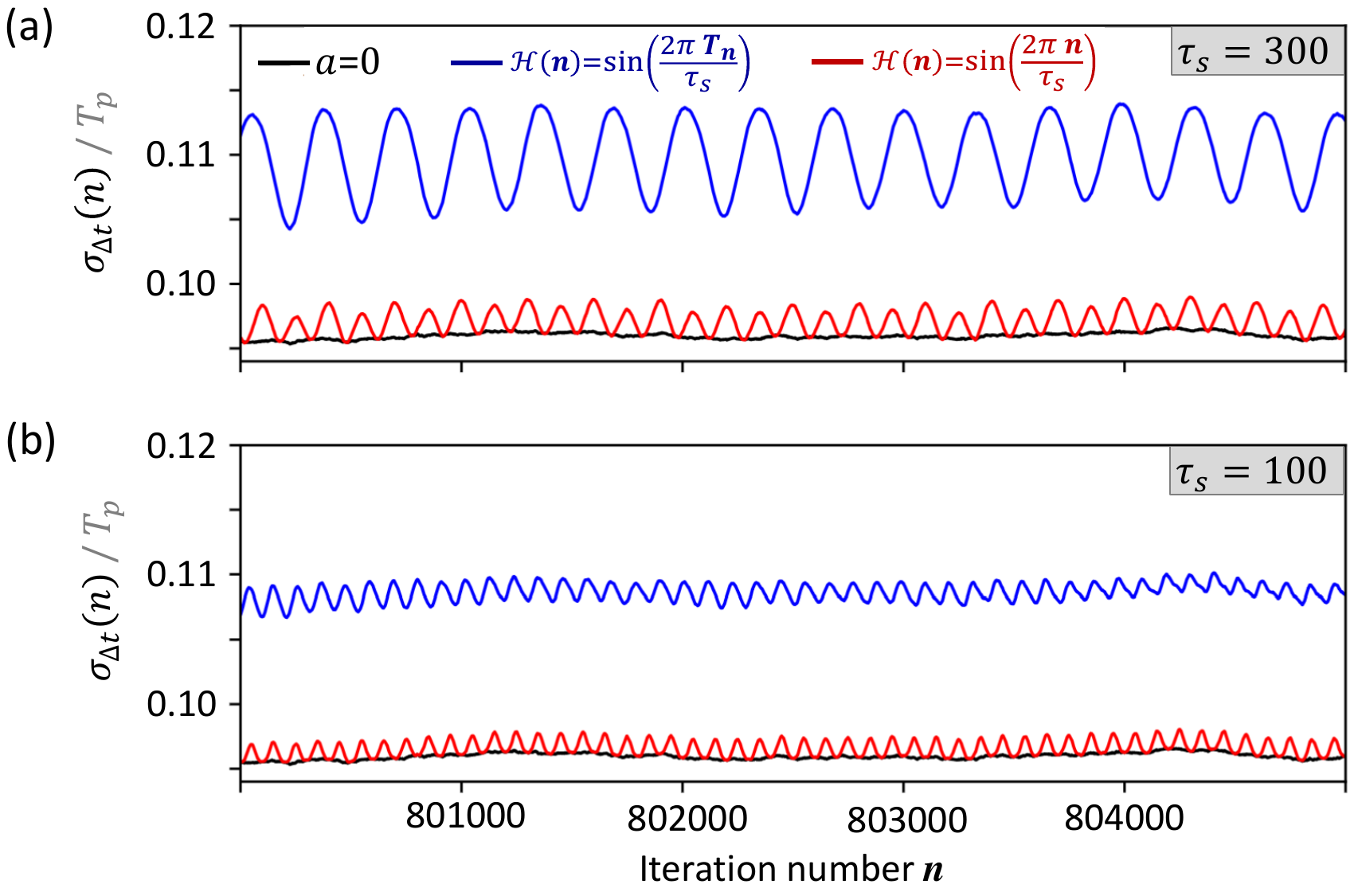}
        \caption{\textbf{Jitter in a driven iterative map}: Normalised standard deviation of the pulse arrival times $\sigma_{\Delta t}(n)$ as a function of the laser cavity round-trip number as reproduced by the iterative map of equation\,\eqref{map}, (a): $\tau_s$=300, (b): $\tau_S$=100. Black line shows the case of a random walk (no forcing function $\mathcal{H}$), blue and red lines show results with different driving terms $\mathcal{H}$ (see legend).
        Parameters: $T_P=1$.}\label{varmap}
    \end{center}
\end{figure}

To understand how the noise-induced modulations affect the variance of the timing fluctuations it is helpful to consider a simple iterative map describing a random process with an added oscillatory term. We consider 
\begin{equation}
 T_{n+1}=T_n+T_P+\xi(n+1)+a\mathcal{H}. \label{map}
\end{equation}
where $T_P$ is the unperturbed inter pulse distance, $\xi(n)$ is a Gaussian white noise term and $\mathcal{H}$ is a sinusoidal function of either the roundtrip number $n$ or the interspike interval $T_n$, i.e.
$\mathcal{H}( n)=\textrm{sin} \left( \frac{2 \pi n}{\tau_s}\right)$ or $\mathcal{H}( T_n)=\textrm{sin} \left( \frac{ 2 \pi T_n}{\tau_s}\right)$, with amplitude
$a$. If $a=0$ then
\begin{equation}
T_{n}=n T_P+\sum_{i=1}^{n} \xi (i)
\end{equation}
and the series $\{ T_{n}-n T_P \}$ is a random walk. In this case, apart from statistical fluctuations, $\sigma_{\Delta t}(n)$ is constant (figure\,\ref{varmap} black line). If $a$ is non-zero and $\mathcal{H}\left( n\right)=\textrm{sin}\left( \frac{2 \pi n}{\tau_s}\right)$ is used, then the sinusoidal term adds regular modulations onto the variance.  For $n$ equal to integer multiples of half the period of the modulations, $\tau_s$, the modulations cancel and $\sigma_{\Delta t}(n)$ is the same as in the $a=0$ case (compare red and black line in figure\,\ref{varmap} where the minima of the red line are on the black curve). If instead $\mathcal{H}\left( T_n\right)=\textrm{sin}\left( \frac{ 2 \pi T_n}{\tau_s}\right)$ is used, the added modulations do not cancel due to the fluctuations in the $T_n$ values that enter the sine function, this leads to an overall increase in the variance (figure\,\ref{varmap} blue line). The mode-locked laser with feedback is comparable to the latter case as the noise induced modulations are not perfectly periodic.

\subsection{Optimized timing jitter reduction via dual-cavity feedback}\label{sec:DFB}

In the previous section we have shown that noise-induced modulations are detrimental to the regularity of the pulse trains produced by passively mode-locked lasers. In this section we investigate the influence of a second feedback cavity on the suppression of the noise induced modulations, and hence on the improvement of the timing regularity of the mode-locked laser output.

Noise-induced modulations are excitations of eigenmodes of a system which are only weakly damped. To ascertain the relative magnitude of the modulations one looks at the eigenvalues of the underlying deterministic system. It has been shown that for oscillatory systems with delay, as e.g. the mode-locked laser subject to resonant feedback from two external cavities, the eigenvalues $\lambda$  can be estimated by the solution of the characteristic equation given in equation\,\eqref{characteristic4}  as a function of the two resonant delay times $\tau_1$ and $\tau_2$ \cite{JAU16a}.

\begin{equation}
\lambda=-(K_1^{\mathrm{eff}}+K_2^{\mathrm{eff}})+K_1^{\mathrm{eff}}e^{-\lambda \tau_1}+K_2^{\mathrm{eff}}e^{-\lambda \tau_2}\label{characteristic4}.
\end{equation}

\begin{figure}[t]
    \begin{center}
        \includegraphics[width=0.9\textwidth]{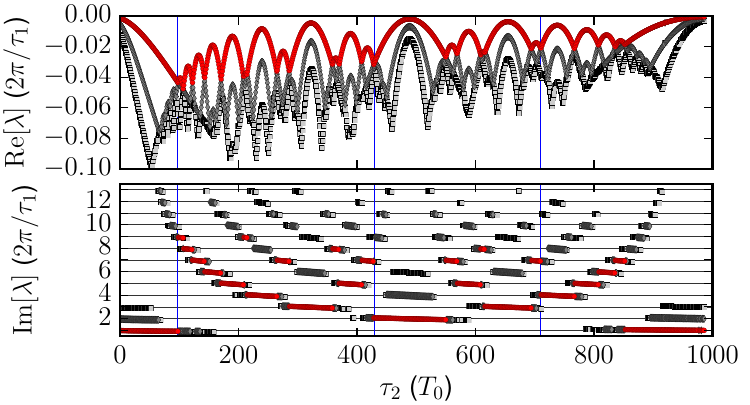}
        \caption{Real and imaginary parts of the largest eigenmodes of a passively mode-locked laser subject to feedback from two external cavities 
        predicted from the fitted characteristic equation equation\,\eqref{characteristic4} \cite{JAU16a}.
        Parameters: $K_1^{\mathrm{eff}}=K_2^{\mathrm{eff}}=0.047$, $\tau_1=1000T_0$.}\label{eigenmodes4}
    \end{center}
\end{figure}

Differences between different oscillatory systems only occur in how strongly the feedback acts upon the system (via the effective feedback strengths  $K_{1,2}^{\mathrm{eff}}$), but not in the form of the characteristic equation \cite{JAU16a}. Therefore, $K_1^{\mathrm{eff}}$ and $K_2^{\mathrm{eff}}$ can be determined numerically via fitting results for small delays to equation \eqref{characteristic4}.  We are interested in the influence of the length of the second feedback cavity, hence, in figure\,\ref{eigenmodes4} the real and imaginary parts of the three largest eigenvalues are plotted as a function of $\tau_2$, for $\tau_1=1000T_0$. The real part gives the damping rate of perturbations and the imaginary part gives the frequency of modulations that are excited by perturbations. This means that the smaller the real parts of the eigenvalues, the smaller the amplitude of the noise-induced modulations will be. For the case at hand
the largest damping rate occurs for $\tau_2=97T_0$.

\begin{figure}[ht]
\centering\includegraphics[width=0.98\columnwidth]{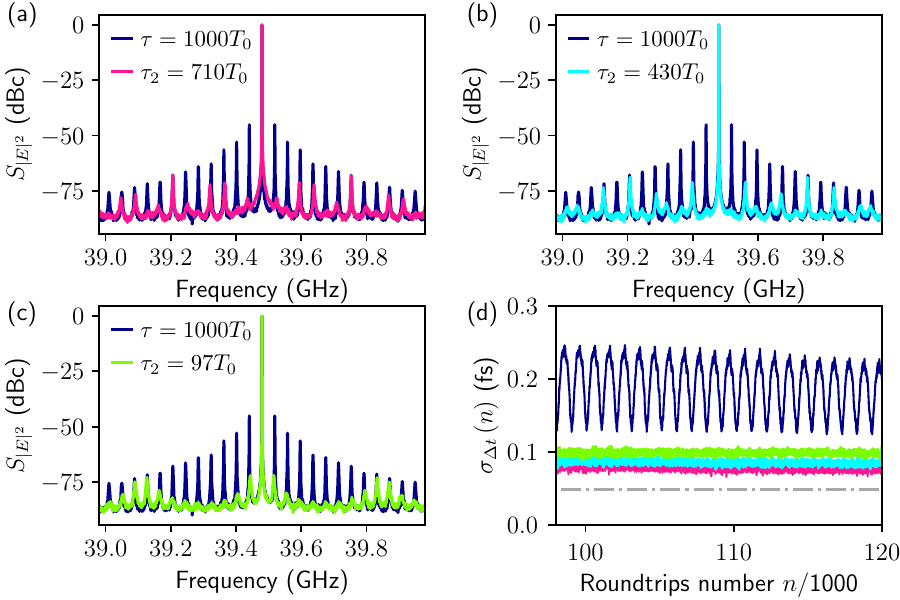}
\caption{(a)-(c) Power spectra of the electric field $S_{|\mathcal{E}|}$ and timing fluctuations $S_{|\Delta t_n|}$ of a passively mode-locked laser subject to feedback from two external cavities for delays indicated by the vertical blue lines in figure\ref{eigenmodes4}. Blue lines show  spectra for the $\tau=1000T_{0}$ single feedback cavity case while overlayed colored spectra result from two delays. (d) Timing jitter $\sigma_{\Delta t}$ (equation \eqref{eq:tj}) for the case of one delay (blue) and two delays (colors as in the spectra in (a)-(c)), grey dashed line shows the long term timing jitter. Parameters: $K=0.1$, $K_1=K_2=0.05$, $\tau_1=1000T_0$, all other parameters as in Table \ref{tab:fonts}.}\label{twofb}
\end{figure}

In figure\,\ref{twofb} power spectra of the electric field amplitude are shown for the three $\tau_2$ values marked by the blue vertical lines
in figure\,\ref{eigenmodes4}. These spectra show a significant suppression of the noise-induced modulations with respect to the single feedback cavity case (dark blue line). As predicted from the eigenvalues, the greatest suppression is achieved for $\tau_2=97T_0$. In addition to the reduction in the amplitude of the noise-induced modulations (indicated by the lower power in the side peaks of the power spectra), the  dominant frequency of the modulations is also changed. 

In order to make a fair comparison to the single feedback cavity case, the total feedback strength is kept the same, i.e. $K_1+K_2=K$. This means that there is a trade off between oscillation suppression and long-term jitter reduction. The semi-analytic analysis in \cite{JAU15} showed that in the absence of noise-induced modulations the long-term timing jitter decreases with increasing resonant feedback lengths according to equation\,\eqref{semi}. Therefore, if $\tau_2$ is shorter the contribution to this effect is reduced. However, the greatest oscillation suppression is expected for a relatively short second feedback cavity and the analysis of the previous section has shown that the noise-induced modulations cause the long-term jitter to increase. To determine the relative influence of these competing effects $\sigma_{\Delta t} (n)$ is plotted in figure\,\ref{twofb}d for the single feedback case (dark blue) and the three dual feedback cases corresponding to figure\,\ref{twofb}a-c (colored). In all three dual feedback cases it can clearly be seen that the modulations are suppressed. The long-term timing jitter is slightly different in the three cases, but in each case it is lower than in the single feedback system. The best results are achieved for $\tau_2=710T_0$. For this feedback configuration the long-term timing jitter is reduced by $\approx$40\% compared with the single feedback case, and the modulation of the timing fluctuations is effectively suppressed. Although the change in the long-term jitter is not very large, the reduction in the modulation amplitude represents a significant improvement in the regularity of the pulse train. For larger $\tau_2$ values the increase in the amplitude of the noise-induced modulations out weighs the influence of the longer cavity and the long-term timing jitter increases.

Experimentally dual feedback configurations have been implemented and, although the delay lengths were not optimised these studies showed a significant improvement in the rms timing jitter \cite{ASG18,JAU16,DRZ13a,HAJ12}. In these works the reduction in the rms timing jitter was larger than the reduction in the long-term jitter demonstrated numerical here. This is expected due to the delay lengths used in these experimental studies. In \cite{JAU16} the delay length of the longer feedback cavity is about 3000 times the interspike interval time. For longer delay times the amplitude of the noise-induced modulations is increased, which causes a greater increase in the variance of the timing fluctuations and hence in the rms and long-term jitter, meaning that comparatively a greater improvement can be achieved with a dual feedback setup.

\section{Discussion and conclusions}

We have shown how noise-induced modulation of the interspike intervals affects the timing regularity of passively mode-locked lasers subject to optical feedback. Although the occurrence of noise-induced modulations in such systems has been observed experimentally \cite{BRE10,SOL93,ARS13}, so far their influence has been neglected or has not been fully taken into account. The results presented here show that this can lead to large over estimates of the regularity of the emitted pulse trains. This is because the long-term timing jitter only gives an indication of the regularity of pulse trains over very long times, and contains no information on fluctuations on shorter time scales. When the delay lengths are very long, then noise-induced fluctuations, which occur on the time scale of the feedback delay time, can cause larger variations in the pulse arrival times than the underlying random-walk-like drift. This effect becomes more important the longer the feedback cavity, as the damping rate of the noise-induced modulations decreases.

A significant improvement in the regularity of the emitted pulse trains can be achieved by adding a second feedback cavity of the appropriate length. This improvement occurs due to the suppression of the noise-induced modulations caused by an increase in the damping rate of the largest eigenmodes of the system. To find the optimal length for the second feedback cavity, minima in the largest eigenvalue need to be found. The eigenvalues depend on the feedback delay times and feedback strengths and as such the minima are not given by fixed ratios of $\tau_1$ and $\tau_2$ \cite{JAU16a}. Generally, though, due to the trade off between oscillation suppression and the increased long-term timing jitter reduction achieved for longer delay times \cite{JAU15}, the second delay length should be about three quarters the length of the first cavity.

By implementing a dual feedback configuration the timing stability of passively mode-locked laser can be substantially improved, making them promising devices for a wide range of applications.



\bibliographystyle{srtnumbered}


\end{document}